\documentclass[reprint]{revtex4-1}

\usepackage{graphicx}% Include figure files
\usepackage{dcolumn}% Align table columns on decimal point
\usepackage{bm}% bold math

\usepackage{amsmath}

\usepackage{geometry}

\usepackage{xcolor}

\begin{document}

\title{Efficient method to create superoscillations with generic target behavior}



\author{Barbara \v{S}oda}

\email{bsoda@pitp.ca} 

\author{Achim Kempf}

\email{akempf@pitp.ca} 

\address{Dept. of Applied Mathematics and Dept. of Physics, University of Waterloo and Perimeter Institute for Theoretical Physics, Waterloo, Ontario, Canada}

\date{\today}

\begin{abstract}
We introduce a new numerically stable meth\-od for constructing superoscillatory wave forms in an arbitrary number of dimensions. The method allows the construction of superoscillatory square-integrable functions that match any desired smooth behavior in their superoscillatory region to arbitrary accuracy. 

%\keywords{Superoscillations \and Superresolution \and Fourier}
\end{abstract}

\maketitle

\section{Introduction}
Superoscillatory functions are functions that locally oscillate faster than the fastest Fourier component that they contain. Superoscillations were first named and made the subject of investigation in work that includes, in particular  \cite{earliest-1}-\cite{beethoven}.   
In hindsight, examples of superoscillatory behavior had previously been seen experimentally, e.g., in optics, and also in theory, where, for example, certain so-called prolate spheroidal functions are superoscillatory. 

In the meantime, both the theory, see, e.g., \cite{theory-1}-\cite{theory-book}, and applications, see, e.g., \cite{apps-1}-\cite{apps-l}, of superoscillations have been actively explored.  

Functions with superoscillatory behavior are the subject of intense investigation for multiple reasons. One reason is that they offer an opportunity for superresolution, i.e., for resolution beyond the diffraction limit. Superresolution is of particular interest in the presence of natural bandwidth limitations, such as the ultraviolet cutoff expected to arise at the Planck scale, see, e.g., \cite{ak-1}-\cite{ak-2}, in the presence of practical bandwidth limitations, for example, arising from absorption bands in media, e.g., in the case of radar \cite{PPA,roadmap,london}, or in the presence of imposed bandwidth limitations, e.g., to avoid deleterious ionization in biological tissue, e.g., in optogenetics \cite{opto-1}-\cite{opto-b}, see \cite{roadmap,london}. 

In certain media, the use of superoscillatory waves for superresolution is obstructed because of the existence of dissipative processes that are faster than the superoscillations. Even in these circumstances, the use of superoscillatory waves can be useful, for example, to explore these fast dissipative processes, \cite{roadmap,london}. 

Also in information theory, see, e.g., \cite{thomascover}, the r\^ole of superoscillatory behavior in communication signals is of interest. There, the study of superoscillatory signals has the potential, for example, for a noise-model independent generalization of the Shannon Hartley theorem \cite{roadmap,london}. 

In quantum theory, superoscillatory behavior has been described early on in the context of weak measurement theory. The further exploration of the quantum mechanics as well as the quantum thermodynamics of superoscillatory behavior is of high interest, see, e.g., \cite{calder}-\cite{angus2},\cite{theory-book},\cite{london}. 

Superoscillatory behavior is known to come with a cost, however, which is that the superoscillatory part of a function necessarily possesses relatively small amplitudes compared to the amplitudes in the rest of the function. In particular, it was proven in \cite{theory-1}-\cite{ak-fe2} that the superoscillatory behavior obeys two scaling laws: for fixed $L^2$-norm, the amplitudes in the superoscillating region necessarily diminish polynomially with the frequency of the imposed superoscillations and they necessarily diminish exponentially with the number of imposed superoscillations. 
\newpage
\section{Motivation} 
The focus of the present paper is the question of how superoscillatory functions can be constructed. At first, it had appeared that superoscillatory functions with numerous fast superoscillations are hard to construct, in the sense that the calculation is numerically unstable. For example, in \cite{beethoven}, a method was presented that allows one, for any fixed $\Omega>0$, to construct $\Omega$-bandlimited and square integrable functions that pass through any finite number of prescribed points. These points can be chosen arbitrarily. If they are chosen such as to force the constructed function to possess superoscillations then the calculation involves the need to invert a matrix that is ill-conditioned. The matrix becomes the more ill-conditioned and, therefore,  the harder to invert numerically, the more superoscillations are being prescribed and the smaller their wavelength. 

Fortunately, however, numerical instabilities in the construction of superoscillatory functions can be avoid\-ed, as was first shown in \cite{leilee}. In prior approaches, such as \cite{beethoven}, the superoscillatory function was constructed by adding a number of non-superoscillatory $\Omega$-bandlimited functions with coefficients that had to be extremely fine-tuned in order to achieve superoscillatory behavior for the resulting sum of these functions, hence the numerical instability. Instead, the method in \cite{leilee} constructs a superoscillatory function as the product of non-superoscillato\-ry functions whose bandwidth adds up to $\Omega$ (recall that as functions are multiplied, their bandwidths add). The multiplicative method of \cite{leilee} allows one to construct superoscillatory functions with any number of arbitrarily close-by zero crossings because the function function inherits the zeros of each of its factors - and these factors can be chosen arbitrarily close to another. The method is numerically stable because no fine tuning is needed to obtain the product function. This method was then used in the later paper \cite{mansuripur} to approximate sinusoidal superoscillations. 

In the present paper, we also build on \cite{leilee}. We obtain a numerically stable method for constructing superoscillatory functions whose superoscillatory stretch exhibits any desired behavior to any desired accuracy, in any number of dimensions. 

\section{New method to efficiently and numerically stably generate superoscillatory functions with arbitrary target behavior in any number of dimensions} 

Let us consider the problem of finding a square integrable function $g: {\rm I\!R}^d \rightarrow {\rm I\!R}$ of a given bandlimit $\Omega$ which, in some given region $B\subset {\rm I\!R}^d$, is approximating some arbitrary given smooth target function $f: {\rm I\!R}^d \rightarrow {\rm I\!R}$ while meeting an arbitrarily given accuracy goal $\epsilon>0$:
\begin{equation}
    |g(x)-f(x)| < \epsilon ~~~\forall x\in B
\end{equation}
Since there is no assumption of bandlimitation for $f$, the behavior of $f$ in $B$ can be chosen arbitrarily highly oscillatory, i.e., the to-be-constructed $\Omega-$bandlimited function $g$ may have to be arbitrarily highly superoscillatory. 

We now show how such a function $g$, that approximates the behavior of $f$ in $B$ to arbitrary given accuracy, can be generated efficiently and numerically stably.   

The idea is to construct $g$ as the product of two functions, one of bandlimit zero that ensures accuracy, namely a Taylor polynomial of the target function $f$, and one function of bandlimit $\Omega$ that ensures square integrability without spoiling the accuracy of the approximation.  

Concretely, the method consists of first Taylor expanding the target function $f$ around some point $p\in B$ to a finite order $n$. 
By Taylor's theorem, we can always choose $n$ large enough so that the Taylor polynomial $f_n$ approximates $f$ in $B$ to meet any given accuracy goal $\epsilon_1>0$: 
\begin{equation}
    |f_n(x)-f(x)|<\epsilon_1 ~~~~~\forall ~ x\in B
\end{equation}
We note that $f_n$ possesses vanishing bandwidth because the Fourier transform of a polynomial is a linear combination of Dirac deltas and their derivatives. We also note that $f_n$ diverges polynomially and that it is therefore not square integrable. 

Second, we therefore also construct an $\Omega$-bandlimited function $c: {\rm I\!R}^d \rightarrow {\rm I\!R}$ which obeys two conditions: 

\begin{enumerate}
\item In the region $B$, we require the function $c$ to stay close to 1 up to some accuracy goal $\epsilon_2>0$:
\begin{equation}
    |c(x)-1| < \epsilon_2~\forall x\in B
\end{equation}
\item We require the function $c$ to decay to zero for $|x|\rightarrow\infty$ sufficiently fast so that  $g(x):=f_n(x)c(x)$ is square integrable in ${{\rm I\!R}^d}$.
\end{enumerate}
For any $\epsilon_2$, such a function $c$ can be  obtained as the product $c(x) = \prod_{i=1}^m c_i(x)$ of a sufficiently large number, $m$, of $\Omega/m$-bandlimited functions, $c_i$, that each take a local maximum of value $1$ at the Taylor expansion point, $p$. Condition (1) is obeyed by choosing $m$ sufficiently large because $c(x)$ becomes increasingly flat around $p$. To see this, we notice that since $c$ takes a local maximum value of $1$ at $p$, its first derivative vanishes, and its second derivative decays with increasing $m$. For example, in one dimension, $(r(x/m)^m)''|_p=r''(x/m)|_p/m$ since $r'(x/m)=0|_p$. In addition, condition 2 is obeyed by choosing $m>n$. 

Clearly, $g:=cf_n$ is $\Omega$-bandlimited and square integrable. We now show that also any accuracy goal $\epsilon$ can be met. To this end, we can choose $\epsilon_1 := \epsilon/2$. This determines $n$ such that $f_n$ differs from $f$ at most by $\epsilon_1$ in $B$. In $B$, $|f_n|$ takes on a maximum, say $M$ at $x_M\in B$. We then meet the overall accuracy goal by choosing any $\epsilon_2>0$ obeying $M-M(1-\epsilon_2)<\epsilon_1$, i.e., by choosing any $\epsilon_2$ obeying $0<\epsilon_2<\epsilon_1/M=\epsilon/2M$. Finally, the choice of $\epsilon_2$ determines $m$. 

The complexification of the method is straightforward. Also, instead of the Taylor expansion, other polynomial expansions may be used. The function $c$ may be called a canvas function because it is designed to provide the blank backdrop on which the pattern of the target function is to be drawn.

\section{Example: a one-dimensional implementation}
Here, we will demonstrate the new method for $\Omega$-band\-limited functions ${\rm I\!R}\rightarrow {\rm I\!R}$. We require the to-be-con\-struc\-ted superoscillatory function, $g$, to approximate the behavior of a target function, $f$, in the interval $B:=[-1,1]$. To this end, we take the MacLaurin series approximation of $f$
\begin{equation}
    f_n(x):= \sum_{k=0}^{n-1} \frac{f^{(k)}(0)}{k!} x^k\approx f(x).
\end{equation}
to order $n$. By Taylor's theorem, we can always find an $n$ large enough to meet any accuracy goal $\epsilon_1$ in $[-1,1]$. We then obtain the desired superoscillatory function 
\begin{equation}
    g(x):=f_n(x) c(x)
\end{equation}
if we can also construct a suitable $\Omega$-bandlimted canvas function $c$ that stays close to $1$ in $[-1,1]$ and that counteracts the $x^{n-1}$ divergence of $f_n$ for $x\rightarrow \pm \infty$. To construct such a function that decays at least as fast as $\frac{1}{x^n}$, we can make, for example, the ansatz:
\begin{equation}
    c(x)= \textup{sinc}\left(\frac{\Omega x}{m}\right)^{m}
\end{equation}
Using the method of integration by differentiation \cite{intbydiff1,intbydiff2,intbydiff3}, we quickly find the Fourier transform of this function:
\begin{eqnarray} \label{fourier}
\tilde{c}(k) & = & \frac{1}{\sqrt{2\pi}} \int_{-\infty}^{\infty} \textup{sinc}\left(\frac{\Omega x}{m}\right)^m e^{i k x} dx\\
& = & \frac{\sqrt{2 \pi}}{\Omega} \frac{\left(\frac{m}{2}\right)^m}{(m-1)!}\nonumber \times\\
& & \times  \sum_{j=0}^m \binom{m}{j} (-)^{j} T_{\frac{1}{m}}^{m-2j} \theta\left(\frac{k}{\Omega}\right)\left(\frac{k}{\Omega}\right)^{m-1}\nonumber
\end{eqnarray}

Here, $T_a$ is the translation operator $T_a=e^{a \Omega \partial/\partial_k}$. 

The work \cite{mansuripur} which followed up on \cite{leilee} also studied the properties of powers of sinc functions. For our purposes here, we will find it useful to further investigate the peculiar property that, near $0$, the sequence of powers of sinc functions approaches a Gaussian in the limit of large $m$. To show this, we could consider the behavior of the binomial coefficients in the Fourier transform  in Eq.\ref{fourier} to investigate how the Fourier transform approaches a Gaussian. 
Instead, let us work directly in the $x$ domain. We use a defining property of Gaussians, $\mathcal{G}(x):=e^{-x^2/(2\sigma^2)}$, namely that they obey:
\begin{equation}
    \frac{\mathcal{G}'(x)}{\mathcal{G}(x)}=-\frac{x}{\sigma^2},
\end{equation}
In comparison, the canvas function, $c(x)=  \textup{sinc}(\frac{\Omega x}{m})^{m}$ obeys for large $m$:
\begin{equation}
    \frac{c'(x)}{c(x)}=-\frac{\Omega^2}{3}  \frac{ x}{m}-\frac{\Omega^4}{45}\left(\frac{x}{m}\right)^3+O\left(\left(\frac{\Omega x}{m}\right)^5\right).
\end{equation}

Therefore, as long as $x \ll m$, the function $c(x)$ is a good approximation to a Gaussian with variance $\frac{\sqrt{3 m}}{\Omega}$. As desired, for increasing $m$, the function $c(x)$ approximates a wider and wider Gaussian so that its amplitude stays closer and closer to $1$ up to $|x|$ of order $m$.  
Only from about $x=\frac{1}{\Omega}\pi n$, where $c(x)$ first drops to zero, $c(x)$ no longer approximates a Gaussian.

Finally, we also need $c$ to meet the accuracy goal $\epsilon_2$. To this end, we notice that while $c(0)=1$, at the ends of the interval $[-1,1]$, the function $c$ falls to approximately $\textup{sinc}(\frac{\pm \Omega}{n})^n=1-\frac{\Omega^2}{6 n}$, where we can again bound the error, and therefore meet any accuray goal $\epsilon_2$ by using Taylor's theorem. 

\section{Outlook}
The new method presented here allows one to construct, efficiently and  numerically stably, a superoscillatory function that possesses any desired smooth behavior in the superoscillating region. The new method should, therefore, be useful for applications ranging from superresolution and the study of fast dissipative processes in media to applications of superoscillations in communication channels or measurement devices, in particular, in scenarios where the channel capacity is limited chiefly by the bandwidth and not by the availability of a large dynamic range.      

In practice, the new method for constructing custom superoscillatory wave forms is straightforward to apply in any circumstances where the desired superoscillatory waveform can be calculated and then produced by a wave form generator, e.g., in acoustics. However, wave form generators can only generate frequencies up to a certain range and at higher frequencies, such as in the optical range, methods other than wave form generators need to be employed to create superoscillatory waves, see e.g., \cite{zheludev1,zheludev2}. In this case, the present method may, however, be applicable directly, namely if the experimental setup allows the physical multiplication of wave forms. In electronics, the multiplication of signals for the purpose of obtaining a superoscillating signal is relatively straightforward through the use of transistors for multiplication. It would be very interesting to explore, for example, the use of photo transistors for the purpose of creating custom superoscillatory electromagnetic waves in optics. 
$$$$
On behalf of all authors, the corresponding author states that there is no conflict of interest. 

\section{Acknowledgements}
AK and B\v{S} acknowledge useful discussions with Lucien Hardy. AK is supported by a Discovery Grant of the Natural Sciences and Engineering Research Council of Canada (NSERC) and by a Google Faculty Research Award. B\v{S} is supported in part by the Perimeter Institute, which is supported in part by the Government of Canada through the Department of Innovation,
Science and Economic Development Canada and by the
Province of Ontario through the Ministry of Economic
Development, Job Creation and Trade.

%\begin{acknowledgements}
%\end{acknowledgements}



\begin{thebibliography}{}

\bibitem{earliest-1} Bucklew, J.A., Saleh, B.E.A.: Theorem for high-resolution high-contrast image synthesis. J. Opt. Soc. Am. A 2 (8), 1233-1236 (1985)

\bibitem{} 
Aharonov, Y., Popescu, S., Rohrlich, D.: How can an infra-red photon behave as a gamma ray. Tel-Aviv University Preprint, 1847–1890 (1990)

\bibitem{} Aharonov, Y., Anandan, J., Popescu, S., Vaidman, L.: Superpositions of time evolutions of a quantum system and a quantum time-translation machine. Phys. Rev. Lett. 64 (25), 2965-2968 (1990)

\bibitem{} Aharonov, Y., Popescu, S., Rohrlich, D., Vaidman, L.: Measurements, errors, and negative kinetic energy. Phys. Rev. A 48 (6), 4084-4090 (1993)

\bibitem{} Berry, M.V.: Evanescent and real waves in quantum billiards and Gaussian beams. J. Phys. A: Math. Gen. 27 (11), 391-398 (1994)

\bibitem{} Berry, M.: Faster than Fourier. In: Anandan J.A.,  Safko, J. (eds.) Fundamental Problems in Quantum Theory, pp. 55-65. World Scientific, Singapore (1994).

\bibitem{} Qiao, W.: A simple model of Aharonov-Berry's superoscillations. J. Phys. A: Math. Gen 29 (9), 2257-2258 (1996)
\bibitem{beethoven} Kempf, A.: Black holes, bandwidths and Beethoven. J. Math. Phys. 41 (4), 2360-2374 (2000) 

\bibitem{theory-1} Ferreira, P.J.S.G., Kempf, A.: The energy expense of superoscillations. In 11th European Signal Processing Conference, 3-6 ept. 2002, Toulouse, France, IEEE, pp.1-4., Print ISSN: 2219-5491, INSPEC Accession Number: 15024475, URL: https://ieeexplore.ieee.org/abstract/document/7072057  (2002)


\bibitem{ak-fe1} Kempf, A., Ferreira, P.J.S.G.: Unusual properties of superoscillating particles. J. Phys. A: Math. Gen. 37 (50), 12067-12076 (2004)

\bibitem{ak-fe2} Ferreira, P.J.S.G., Kempf, A.: Superoscillations: faster than the Nyquist rate. IEEE transactions on signal processing 54 (10), 3732-3740 (2006)

\bibitem{ak-fe3} Ferreira, P.J.S.G., Kempf, A., Reis, M.J.C.S.: Construction of Aharonov–Berry's superoscillations. J. Phys. A: Math. Theor. 40 (19), 5141–5147 (2007)

\bibitem{calder} Calder, M.S., Kempf, A.: Analysis of superoscillatory wave functions. J. Math. Phys. 46 (1), 012101 (2005)

\bibitem{} Tang, E., Garg, L., Kempf, A.: Scaling properties of superoscillations and the extension to periodic signals. J. Phys. A: Math. Theor. 49, (33), 335202 (2016)

\bibitem{leilee} Chojnacki, L., Kempf, A.: New methods for creating superoscillations. J. Phys. A: Math. Theor. 49 (50), 505203 (2016)

\bibitem{angus1} Prain, A.: Vacuum Energy in Expanding Spacetime and Superoscillation-Induced Resonance. Master's thesis, University of Waterloo.  https://uwspace.uwaterloo.ca/handle/10012/3700 (2008)

\bibitem{angus2} Kempf, A., Prain, A.: Driving quantum systems with superoscillations. J. Math. Phys. 58 (8), 082101 (2017)

\bibitem{} Lee, D.G., Ferreira, P.J.S.G.: Superoscillations with optimal numerical stability. IEEE Signal Processing Letters 21 (12), 1443-1447 (2014)

\bibitem{} Hao, Y., Kempf, A.: On the stability of a generalized Shannon sampling theorem. In Information Theory and Its Applications, 2008. Intl. symposium ISITA 2008. pp. 1-6. IEEE, 2008.

\bibitem{} Berry, M.V., Popescu, S.: Evolution of quantum superoscillations and optical superresolution without evanescent waves. J. Phys. A: Math. Gen. 39 (22), 6965–6977 (2006)

\bibitem{} Aharonov, Y., Colombo, F., Sabadini, I., Struppa, D.C., Tollaksen, J.: Some mathematical properties of superoscillations. J. Phys. A: Math. Theor. 44 (36), 365304  (2011) 

\bibitem{theory-l} Berry, M.V.: Suppression of superoscillations by noise. J. Phys. A: Math. Theor. 50 (2), 025003 (2016)

\bibitem{theory-book} Aharonov, Y., Colombo, F., Sabadini, I., Struppa, D., Tollaksen, J.: The mathematics of superoscillations, Memoirs of the AMS 247 (2017), 1174


\bibitem{apps-1} Dennis, M.R., Hamilton, A.C., Courtial, J.: Superoscillation in speckle patterns. Optics letters 33 (24), 2976-2978 (2008)

\bibitem{} Aharonov, Y., Erez, N., Reznik, B.: Superoscillations and tunneling times. Phys. Rev. A 65 (5), 052124 (2002) 

\bibitem{zheludev1} Huang, F.M., Chen, Y., Garcia de Abajo, F.J., Zheludev, N.I.: Optical super-resolution through super-oscillations. J. Opt. A: Pure Appl. Opt. 9-9, S285–S288 (2007)

\bibitem{zheludev2} Rogers, E.T.F., Zheludev, N.I.: Optical super-oscillations: sub-wavelength light focusing and super-resolution imaging. J. Opt. 15 (9), 094008 (2013)

\bibitem{} Makris, K.G., Psaltis, D.: Superoscillatory diffraction-free beams. Opt. Lett. 36 (22), 4335-4337 (2011)

\bibitem{} Berry, M.V.: A note on superoscillations associated with Bessel beams. J. Opt. 15 (4), 044006 (2013)

\bibitem{} Katzav, E., Schwartz, M.: Yield-optimized superoscillations. IEEE Transactions on Signal Processing 61 (12), 3113-3118 (2013)

\bibitem{} Dubois, M., Bossy, E., Enoch, S., Guenneau, S., Lerosey, G., Sebbah, P.: Time-driven superoscillations with negative refraction. Phys. Rev. Lett. 114 (1), 013902 (2015)

\bibitem{} Hyvärinen, H.J., Rehman, S., Tervo, J., Turunen, J., Sheppard, C.J.R.: Limitations of superoscillation filters in microscopy applications. Opt. Lett. 37 (5) 903-905 (2012)

\bibitem{} Huang, K., Ye, H., Teng, J., Yeo, S.-P., Luk'yanchuk, B., Qiu, C.-W.: Optimization‐free superoscillatory lens using phase and amplitude masks. Laser Photonics Rev. 8 (1), 152-157 (2014)

\bibitem{} Sokolovski, D., Sala Mayato, R.: "Superluminal” transmission via entanglement, superoscillations, and quasi-Dirac distributions. Phys. Rev. A 81 (2), 022105 (2010)

\bibitem{} Baranov, D.G., Vinogradov, A.P., Lisyansky, A.A.: Abrupt Rabi oscillations in a superoscillating electric field. Opt. Lett. 39 (21), 6316-6319 (2014)

\bibitem{} Eliezer, Y., Bahabad, A.: Super-transmission: the delivery of superoscillations through the absorbing resonance of a dielectric medium. Opt. Express 22 (25), 31212-31226 (2014) 

\bibitem{} Zheludev, N.I.: What diffraction limit?. Nat. Mater. 7 (6), 420-422 (2008)

\bibitem{apps-l} Aharonov, Y., Colombo, F., Sabadini, I., Struppa, D.C., Tollaksen, J.: Evolution of superoscillatory data. J. Phys. A: Math. Theor. 47 (20), 205301 (2014)

\bibitem{PPA} Kempf, A.: Signals maximizing information return in penetrating radar, US provisional patent application \#: 61/629,930  (2012)

\bibitem{opto-1} Pastrana, E.: Optogenetics: controlling cell function with light. Nat. Methods 8 (1), 24-25 (2010)

\bibitem{}  Deisseroth, K.: Optogenetics. Nat. Methods 8 (1), 26-29 (2011)

\bibitem{} Editorial: ``Method of the Year 2010". Nature Methods. 8 (1), (2010)

\bibitem{opto-l} News, S.: Insights of the decade. Stepping away from the trees for a look at the forest. Introduction. Science 330, 1612-1613 (2010)

\bibitem{opto-a} Jimenez, J.C., Su, K., Goldberg, A.R., Luna,  V.M., Biane, J.S., Ordek, G., Zhou, P. et al.: Anxiety Cells in a Hippocampal-Hypothalamic Circuit. Neuron. 97 (3), 670-683 (2018)

\bibitem{opto-b} Chen, S., Weitemier, A.Z., Zeng, X., He, L., Wang, X., Tao, Y., Huang, A.J.Y., Hashimotodani, Y., Kano, M., Iwasaki, H., Parajuli, L.K., Okabe, S., Loong Teh, D.B., All, A.H., Tsutsui-Kimura, I., Tanaka, K.F.,  Liu, X., McHugh, T.J.: Near-infrared deep brain stimulation via upconversion nanoparticle–mediated optogenetics. Science 359 (6376), 679-684 (2018)  

\bibitem{thomascover} Cover, T.M., Thomas, J.A.: Elements of information theory. 2nd edition. Wiley, Hoboken, New Jersey (2006)

\bibitem{ak-1} Kempf, A.: Fields over unsharp coordinates. Phys. Rev. Lett. 85 (14), 2873-2876 (2000)

\bibitem{} Kempf, A.: Covariant information-density cutoff in curved space-time. Phys. Rev. Lett. 92 (22), 221301 (2004)


\bibitem{} Hao, Y., Kempf, A.: On a non-Fourier generalization of Shannon sampling theory.  10th Canadian Workshop on Information Theory (CWIT),  Edmonton, Canada,  6-8 June 2007. IEEE, pp. 193-196 (2007)


\bibitem{} Kempf, A.: Information-theoretic natural ultraviolet cutoff for spacetime. Phys. Rev. Lett. 103 (23), 231301 (2009)

\bibitem{} Kempf, A.: Spacetime could be simultaneously continuous and discrete, in the same way that information can be. New J. Phys. 12 (11), 115001 (2010)

\bibitem{jason} Pye, J., Donnelly, W., Kempf, A.: Locality and entanglement in bandlimited quantum field theory. Phys. Rev. D 92 (10), 105022 (2015)

\bibitem{ak-2} Aasen, D., Bhamre, T., Kempf, A.: Shape from sound: toward new tools for quantum gravity. Phys. Rev. Lett. 110 (12), 121301 (2013)

\bibitem{mansuripur} Mansuripur, M., Jakobsen, P.K.: An approach to constructing super oscillatory functions. J. Phys. A: Math. Theor. 52 (30), 305202 (2019)

\bibitem{roadmap} Berry, M.V., Kempf, A., et al: Roadmap on Superoscillations. J. of Optics 21, 053002 (2019) 

\bibitem{london} Kempf, A.: Four aspects of superoscillations, Quantum Studies: Mathematics and Foundations. 5, 477-484 (2018)

\bibitem{intbydiff1} Kempf, A., Jackson, D.M., Morales, A.H.: New Dirac delta function based methods with applications to perturbative expansions in quantum field theory, J. Phys. A: Math. Theor. 47 (41) 415204 (2014)

\bibitem{intbydiff2} Kempf, A., Jackson, D.M., Morales, A.H.: How to (path-) integrate by differentiating, J. Phys.: Conference Series  626 (1)  012015 (2015)

\bibitem{intbydiff3} Jia, D., Tang, E., Kempf, A.: Integration by differentiation: new proofs, methods and examples. J. Phys. A: Math. Theor. 50 (23), 235201 (2017)



% etc
\end{thebibliography}


\onecolumngrid{

}

\end{document}